\documentstyle[epsf,floats,multicol,prb,aps]{revtex} 

\input epsf.sty  
\begin{document}  
\draft  
\title{Magnetic Field scaling of Relaxation curves in Small Particle Systems}  
\author{\`Oscar Iglesias and Am\'{\i}lcar Labarta}  
\address{Department de F\'{\i}sica Fonamental, Facultat de F\'{\i}sica, Universitat\\  
de Barcelona, Diagonal 647, 08028 Barcelona, Spain}  
\date{Printed on:\today, Last version: 28/05/2001}  
\maketitle  
\newcommand{\svar}{T \ln(t/\tau_0)}  
\newcommand{\phrb}{Phys. Rev. B }  
\newcommand{\phrl}{Phys. Rev. Lett. }  
\newcommand{\phre}{Phys. Rev. E }  
\newcommand{\jm}{J. Magn. Magn. Mater. }  
\newcommand{\psa}{Phys. Stat. Sol. A }   
\newcommand{\jap}{J. Appl. Phys. }   
\newcommand{\beq}{\begin{equation}} 
\newcommand{\eeq}{\end{equation}} 
  
\begin{abstract}  
  
We study the effects of the magnetic field on the relaxation of the magnetization of small monodomain non-interacting particles with random orientations and distribution of anisotropy constants. 
Starting from a master equation, we build up an expression for the time dependence of the magnetization which takes into account thermal activation only over barriers separating energy minima, which, in our model, can be computed exactly from analytical expressions. Numerical calculations of the relaxation curves for different distribution widths, and under different magnetic fields H and temperatures T, have been performed. We show how a $\svar$ scaling of the curves, at different T and for a given H, can be carried out after proper normalization of the data to the equilibrium magnetization. The resulting master curves are shown to be closely related to what we call effective energy barrier distributions, which, in our model, can be computed exactly from analytical expressions. The concept of effective distribution serves us as a basis for finding a scaling variable to scale relaxation curves at different H and a given T, thus showing that the field dependence of energy barriers can be also extracted from relaxation measurements.
\end{abstract}  

     
  
 
\begin{multicols}{2} 
\section{Introduction}  
  
\label{introduction}  
Time dependent phenomena in small-particle systems have been the subject of  
an increasing number of experiments because of their interest as  
non-equilibrium phenomena in spin systems, \cite{Dormannadv97} for  
magnetic recording materials technology \cite{JMMM221} and even as a possible way to prove  
experimentally the existence of macroscopic quantum tunneling phenomena in  
magnetic materials. \cite{StampCB,Chichilpr} 
Whereas the basis of a theory of the magnetic  
after-effect dates back from old studies on rock magnetism, \cite{Stoner48,Neel2,Street} the 
interpretation of several experimental results is still waiting for suitable  
theoretical models that capture the relevant factors and parameters that can  
play a role in the explanation of these phenomena. One of the points that  
has not been completely clarified is the influence of a magnetic field in  
the relaxation of small-particle systems.  
  
Relaxation in zero field is usually analyzed in terms of parameters such as  
the so-called magnetic viscosity $S$, \cite{Chantrelljm91} fluctuation field \cite{Wohlfarthjpf84,Chantrellpss86,deWittejm90} and activation volume,  \cite{Givordjm87,Givordjm88} which are susceptible to misinterpretations.  
In the last years, several authors \cite{Labartaprb93,Iglesiasjm95,Iglesiaszpb96,Balcellsprb97,Barbarajm93,Vincentjpe94,Wernsdorferjm95} have proposed an alternative method to  
analyze relaxation curves based on a $T\ln (t/\tau _0)$ scaling of the  
relaxation data at different temperatures that avoids the above mentioned  
problems and gains insight on the microscopic details of the energy barrier  
distribution $f(E)$ producing the relaxation. \cite{Iglesiaszpb96,Balcellsprb97} In this context, the  
purpose of this article is to extend this kind of analysis to the case of  
relaxation in the presence of a magnetic field. We want to account for the  
experimental studies on the relaxation of small-particle systems, which  
essentially measure the acquisition of  
magnetization of an initially demagnetized sample under the application of a  
magnetic field. \cite{Vincentjpe94,Kimjm99,Lisfijm99,Mayergoyzjap99,Montseprb99} In this kind of experiments, the field  
modifies the energy barriers of the system that are responsible for the time  
variation of the magnetization, as well as the final state of equilibrium 
towards which the system relaxes.
The fact that usually the magnetic  
properties of the particles (anisotropy constants, easy-axis directions and  
volumes) are not uniform in real samples, adds some difficulties to this  
analysis because the effect of the magnetic field depends on them in a  
complicated fashion. In a previous study, \cite{Labartaprb93,Balcellsprb97} we started to address some  
of these peculiarities, showing how experimental relaxation data must be  
treated in order to compare relaxation curves at different temperatures  
and fields making simple assumptions about the sample composition. Here,  
we will present  
the theoretical background that supports this phenomenological approach, 
as well as detailed numerical calculations of the time  
dependence of the magnetization of a system of non-interacting randomly  
oriented small monodomain particles with uniaxial anisotropy and with a  
distribution of anisotropy constants. In a first approximation, we will  
neglect inter-particle interactions leaving for a future investigation the  
effects of long-ranged dipolar interactions between the particles.  
  
The paper is organized as follows. In Sec. \ref{model}, we present the basic  
features of model show how the distribution of energy barriers of the system is influenced b the application of a magnetic field with the help of the concept of effective energy barrier distribution. 
In Sec. \ref{2state} we introduce the Two-State Approximation (TSA) 
for the calculation of the thermal dependence of the equilibrium magnetization
In Sec. \ref{relaxation},  
we derive the equation governing the time dependence of the magnetization  
from a master rate equation in the TSA. The results of numerical calculations based on the above mentioned equation are presented in Sec.\ref {numerical}. 
There, we present the $\svar$ scaling of relaxation curves at a given magnetic field, discussing its range of validity. We also study the possibility of a scaling 
at different fields and fixed temperature, and its applications.
Finally in Sec. \ref{conclusions} we resume the main conclusions of the  
article.  
  
\section{Model}  
\label{model}  
 
We consider an ensemble of randomly oriented noninteracting single-domain ferromagnetic particles of volume $V$ and magnetic moment ${\bf M}={M_s}V\bf m$ with uniaxial anisotropy.  
To take into account the spread of particle volumes in real  
samples, we will assume that the particles anisotropy constants $K$ are  
distributed according to some function $f(K)$. 

\begin{figure} 
\epsfxsize=8.6cm 
\leavevmode 
\epsfbox{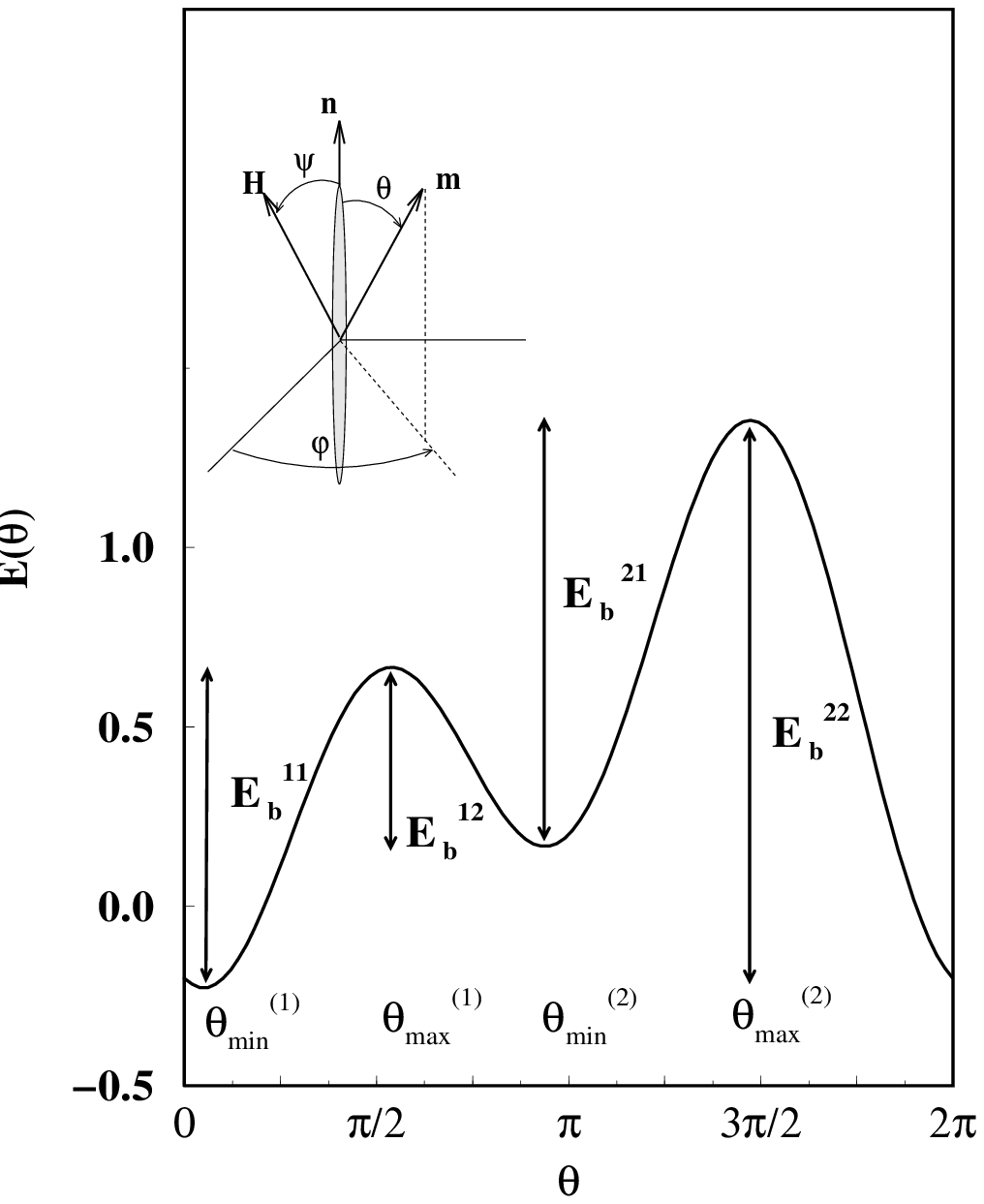} 
\caption{Energy function $E(\theta,\psi)$ as a function of the angle between 
the magnetization vector $\bf m$ and the magnetic field $\bf h$, for $\bf m$ 
in the plane of the easy-axis $\varphi=0$,
as given by Eq. (\ref{energy2}).  
The plot is for a particle whose easy-axis $\bf n$ forms an angle $\psi= 30^o$ 
with $\bf h$, and $H=0.3$. We have used the following  
notation to designate the extrema of the energy: 
$\theta^1_{\rm min}$ and $\theta^1_{\rm max}$ refer to the extrema closer to the 
field direction while $\theta^2_{\rm min}$ and $\theta^2_{\rm max}$ refer to  
those further from the direction of the field. The four possible  
energy barriers between them are $E^{ij}_b\equiv E(\theta^i_{\rm max})- 
E(\theta^j_{\rm min})$.
Inset: Schematic representation of the quantities involved in the  
definition of the system. The easy-axis of the particles $\bf n$ are in
the x-z plane forming an angle $\psi$ with the magnetic field $\bf H$, 
which points along the z axis. $\theta$ and $\varphi$ are the spherical 
angles of the magnetization vector $\bf M$.  
} 
\label{energyplot} 
\end{figure}

The energy of a particle is determined by the orientation of 
$\bf M$ with respect to the external magnetic field  
$\bf H$ and to the easy-axis direction $\bf n$. Using the angular 
coordinates defined in Fig. \ref{energyplot}, it can be written as 
\begin{equation} 
\label{energy2} 
\bar E=\frac{E}{V K}=-\cos^2(\theta)-2h\cos(\theta-\psi) \ .
\end{equation}
where we have defined the $\it reduced$ $\it field$ $h\equiv H/H_{\rm c}$  
and $H_{\rm c}=2K/M_s$ as the $\it critical$ $\it field$ for an aligned particle.
We have concentrated in the two dimensional case
($\bf M$ lying in the plane formed by $\bf H$ and $\bf n$; $\varphi=0$,
since the energy maxima and minima can be calculated analytically 
only in this case.
In Fig. \ref{energyplot}, we show the variation of the energy with $\theta$
for a typical case, defining in the same figure the notation for the energy barriers and extrema.  

\subsection{Effective energy barrier distribution}  
\label{effbar}  
 
The magnetic field  
modifies the energy barriers of the system depending on the  
particle orientation and anisotropy value, and, consequently, changes the original energy barrier distribution. \cite{Berkovjm92} 
Let $E_{\rm b}^0$ be the energy barrier in zero field.  
Then, for a particle oriented at an angle $\psi$, $h$ modifies the barrier  
by a factor $g(h,\psi )$ in the following form \cite{Berkovjm92,Victoraprl91}
\begin{equation}  
\label{efield} 
E_{\rm b}=E_{\rm b}^0\ g(h,\psi)\ .   
\end{equation}  
If $f(E_{\rm b}^0)$ is the energy barrier distribution in zero field, which has  
in fact the same functional dependence than the distribution of anisotropy constants $f(K)$, then the distribution in the presence of a field is simply modified to   
\begin{equation}  
\label{feff0}  
f_{\rm eff}(h,E_{\rm b},\psi )=f(E_{\rm b}^0)\left( {\frac{\partial E_{\rm b}^0(E_{\rm b})}{\partial E_{\rm b}}}  
\right) =f(E_{\rm b}^0)/g(h,\psi )\ ,   
\end{equation}  
which we will call effective energy barrier distribution.  
 
\begin{figure} 
\epsfxsize=8.6cm 
\leavevmode 
\epsfbox{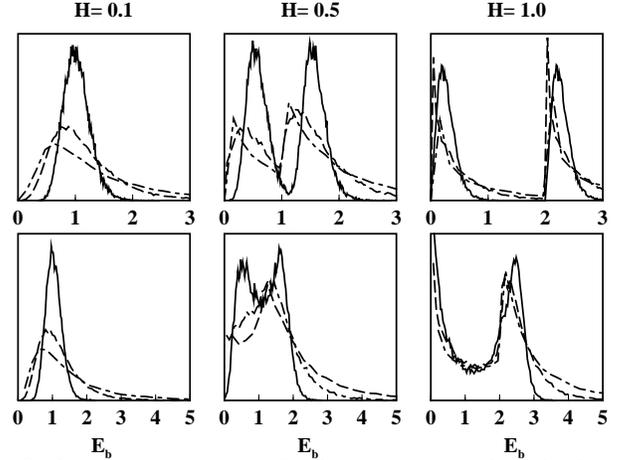} 
\caption{Upper panels: effective energy barrier distributions for aligned particles with a lognormal distribution of anisotropy constants of 
$\sigma= 0.2$ (continuous lines), 0.5 (dashed lines), 
0.8 (dot-dashed lines) 
for values of $H$ as indicated in the figures. 
Lower panels: same as upper panels but for particles with random orientations 
of anisotropy axes.
} 
\label{distriali} 
\end{figure}

In order to understand the qualitative change of $f_{\rm eff}$ with $h$,  
we have numerically calculated $f_{\rm eff}(E_{\rm b})$   
for a system of oriented particles with logarithmic-normal distribution of anisotropies 
\begin{equation}  
\label{logn}  
f(K)={\frac{1 }{{\sqrt{2\pi}K \sigma}}} e^{-  
\ln^2({K/K_0})/{2\sigma^2}}  \ ,
\end{equation}
for different widths $\sigma$ and $K_0=1$, and several values of the 
magnetic field $h$.  
The calculation have been performed by making energy barriers histograms  
for a collection of 10~000 particles.  
The results are given in Fig. \ref{distriali} (upper panels).   
In all the cases, we observe the progressive splitting of the original  
distribution $f(E_{\rm b}^0)$ in two subdistributions of high and low barriers  
as $h$ increases from zero. The field tends to make deeper one of the minima, 
therefore increasing the two energy barriers for rotation of $\bf M$ 
out of the field direction, while the other two are reduced. 
In this way, the global effect of $h$ is a splitting of $f(E_{\rm b})$ towards  
lower and higher values of $E_{\rm b}$.    
 
As $h$ attains the critical value $h_{\rm c}$ for the particles with smaller $K$,  
a peak of zero or almost zero  
energy barriers starts to appear (see for example the curves  
for $H=0.5, 1.0$ in the case $\sigma=0.5$); 
while most of the non-zero barriers are distributed according to a 
distribution identical to $f(E_{\rm b}^0)$, but centered at higher energies.  
The higher the width of the distribution $\sigma$,  
the lower the $h$ at which the lowest energy barriers  
start to be destroyed by the field.  
 
Finally, the combined effect of random orientations and $f(K)$ has been 
considered. The results are shown in Fig. \ref{distriali} (lower panels), 
where we can see that the features of the preceding case are still observed. Now, at high $h$, the distributions are smeared out by the disorder, and the minima 
becomes less pronounced due to the spread in particle orientations. 
 
In Sec. \ref{numerical}, we will discuss how these results affect the time dependence of magnetization in relaxation experiments. 
  
\section{Two-state approximation}  
\label{2state} 
 
The calculation of the equilibrium magnetization at non-zero $T$ and finite $K$ proceeds along the standard techniques of statistical mechanics. For particles oriented at an angle $\psi$, $m(H,T)$ is simply given by the average of the projection of the magnetic moment of the particles onto the field direction over all their possible orientations $\theta$. In our model, this is \cite{Creggjm99,Chantrelljm85}
\begin{equation}  
\label{magn1}  
m(H,T,\psi)={\frac{1 }{{\cal Z}}} \int_{{\bf \Omega}} d{\bf \Omega}  
\cos\theta e^{-U(\theta,\psi)} \ , 
\end{equation}  
where ${\bf \Omega}$ is the solid angle and ${\cal Z}$ is the partition  
function of the system. 
Here, the energy $U(\theta,\psi)$ appearing in the Boltzmann probability,  
has to be calculated from Eq. (\ref{energy2}), then 
\begin{equation}  
\label{U} 
U(\theta)= -\alpha\sin^2\theta+\xi\cos(\theta-\psi)\ , 
\end{equation}  
where the two dimensionless parameters
\begin{equation}  
\alpha \equiv \frac {\mu KV}{k_BT} \ \ , 
\xi \equiv \frac {\mu HV}{k_BT} \ , 
\end{equation}  
have been introduced. 

At $T$ such that the thermal energy $k_BT$ is smaller than the relevant 
energy barriers of the system, typically of the order of the anisotropy energy 
$KV$ ($\alpha\gg 1$), the main contribution to thermodynamic averages comes from states around the energy minima, since thermally activated jumps out of the  
stable directions of the magnetization have extremely low probability to  
succeed. 
Therefore, as it will be useful for the numerical calculations of 
the relaxation curves in Sec. \ref{numerical}, we will consider the so-called 
{\it {Two-State Approximation}} (TSA). \cite{Pfeifferpsa90b,Pfeifferpsa90c} 
In this approximation, the continuum of states corresponding to all the 
possible orientations of $\bf m$ is truncated to the two local energy minima
states.

This will allow us to replace the integrations over magnetization  
directions by sums over the two energy minima. If the particle has only  
one minimum, the two states considered in the calculation will be the minimum  
and the maximum of the energy function.  
For a system of randomly oriented particles and with  
a distribution of anisotropy constants $f(K)$, Eq. (\ref{magn1}) becomes in the TSA 
\begin{equation}  
\label{magnts} 
m_{TS}(H,T)=\int_0^\infty dK\int_0^\pi  
d\psi f(K)\bar m_{TS}(K,\psi ),   
\end{equation}  
where   
\begin{equation}  
\label{magntsalf} 
\bar m_{TS}(K,\psi )={\frac 1{{\cal Z}}}\sum_{i=1,2}\cos [\theta  
_{\rm min}^i(\psi )]e^{-{E_{\rm min}^i(K,\psi )\beta }}   
\end{equation}  
stands for the magnetization of an individual particle in the TSA, 
and $\beta=1/k_B T$.  
  
\begin{figure} 
\epsfxsize=8.6cm 
\leavevmode 
\epsfbox{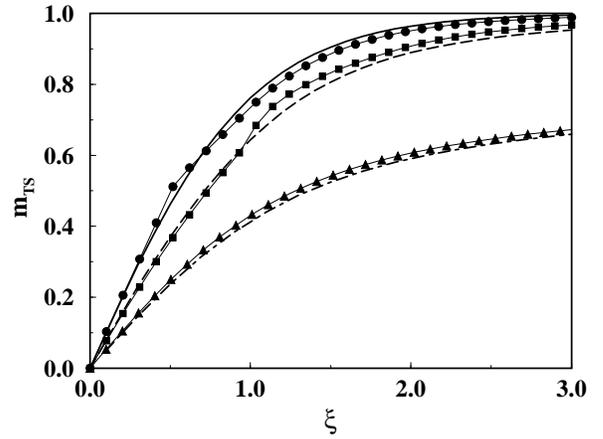} 
\caption{Magnetization curves as a function of the dimesionless Zeeman energy 
$\xi=\mu HV/k_B T$ in the TS approximation. 
Symbols stand for randomly oriented particles with $K_0 = 0.5, 1.0, 10.0$ 
(from the uppermost curve).
The case $K_0 =10$ is compared to the exact 
result given by Eq. (\ref{magn1})(dash-dotted line).
The case $K_0 =1$ is compared with a system of randomly oriented particles 
with $f(K)$, $K_0= 1$ and $\sigma= 0.5$ (long-dashed line)
The result for aligned particles is displayed as a continuous line, 
for which $m_{TS}= \tanh(\xi)$ 
($m_{\rm TS}$ is independent of $\sigma$ in this case).
} 
\label{mtsplot} 
\end{figure}

Eq. (\ref{magnts}) has been numerically evaluated for a system of randomly oriented 
particles and several values of $K_0$ and the results are displayed in Fig. \ref{mtsplot}. For the smallest $K_0$ values, the curves present a small jump 
at a certain value of $\xi$. This may seem unphysical but, in fact, this jump appears at an $h$ equal to the critical field for the disappearance of one of the energy minima. 
In fact, when averaging over a distribution of anisotropies $f(K)$
with $K_0=1$ and $\sigma= 0.5$, this jump disappears.
 
As expected, the TSA curves coincide with the results obtained from the 
exact expression Eq. (\ref{magntsalf}) for high enough $K_0$ (compare the $K_0= 10$
case with the dashed-dotted line in Fig. \ref{mtsplot}). On the other hand, 
at low enough $K_0$, the TSA reproduces the exact result for aligned particles, 
for which the magnetization curve reduces to $m_{\rm TS}= \tanh(\xi)$, since the 
magnetization does not depend on $\alpha$ in this case (compare the continuous line with the case $K_0 =0.5$). \cite{GarciaPaladv00}

\section{Relaxation curves in the presence of a magnetic field}  
\label{relaxation}  
 
Within the context of the Fokker-Planck equation \cite{Brownpr63,Coffeybook} 
for {\bf M} in the discrete 
orientation approximation, \cite{Pfeifferpsa90b,Pfeifferpsa90c}  
we will assume that the relaxation of the magnetization due to thermal fluctuations can be modeled by a markovian stochastic process. Its dynamics can then be described by a master equation for $P_i$, the probability to find the  
magnetization vector at time $t$ in the equilibrium state $i$.  
Furthermore, we will assume that we are in the regime where the TSA is valid  
and, consequently, only transitions between the two  
equilibrium directions of the magnetization given by the minima of the  
energy (\ref{energy2}) will be considered.
Moreover, in models considering continuous variables for the numerical
evaluation of relaxation dynamics \cite{Wolffprl89,Lyberatosjpd00,Nowakann01}, the elementary time step depends on $T$ and $H$, giving rise to relaxation curves which are not directly comparable.
In a recent work, Novak and Chantrell \cite{Nowakprl00} have
faced the problem of the quantification of the time step used in Monte Carlo simulation, giving a method to quantify the time step in real units. As an alternative, we propose a simple dynamical model that avoids this problem since, in the TSA, it can be solved analytically in terms of intrinsic parameters. 
  
Taking into account that the transitions between the  
two minima can take place either by jumping over the barrier placed  
to the right or to the left of the initial state, the master equation governing the time dependence of the  
magnetization can be written as \cite{Reifbook}   
\begin{equation}
\label{difeq} 
\frac{dP_i}{dt}= \sum_{k=1,2}\sum _{j \neq i}  
\left\{ w_{ji}^{(k)}P_j - w_{ij}^{(k)}P_i \right\}  \ ,  
\end{equation}  
where $w_{ij}^{(k)}$ designates the transition rate for a jump  
from the state $i$  
to the state $j$ separated by the maximum $k$ (see Fig. \ref{energyplot}). 
The transition rates can be freely assigned as long as to fulfill  
the detailed balance condition. \cite{Reifbook} 
It is a common choice to consider the Boltzmann probability 
with the energy difference between the two minima in the exponent. This  
choice, in spite of giving the correct thermodynamic averages in a Monte Carlo simulation, may not be appropriate to describe the dynamics of the system, since the energy barriers between the minima are not taken into account.
 
For this reason, in the exponential of the Boltzmann probability,  
we have considered the energy difference between the initial minimum 
$i$ and the maximum $k$ that separates it from the final state $j$ 
\begin{equation}  
w_{ij}^{(k)}=\frac{1}{\tau_{ij}^{(k)}}= 
\frac{1}{\tau_0} e^{-E_{\rm b}^{ki} \beta},   
\end{equation}  
where $\tau_0^{-1}$ is the attempt frequency.  
It is a trivial matter to prove that the following detailed balance equation  
holds   
\begin{eqnarray}  
\frac{w_{21}^{T}}{ w_{12}^{T}}= 
\frac{w_{21}^{(1)}+w_{21}^{(2)}}{w_{12}^{(1)}+w_{12}^{(2)}} = 
e^{-\beta \varepsilon}\ ,  
\end{eqnarray}  
guaranteeing that thermal equilibrium is reached in the long time limit. \cite{Reifbook} 
$\varepsilon = E_{\rm min}^1-E_{\rm min}^2$ is a measure of the  
asymmetry of the energy function.  
 
Taking into account the normalization condition $P_1+P_2=1$,  
one can easily solve Eq. (\ref{difeq}) for $P_1$ and $P_2$ as a function  
of time 
\begin{eqnarray}  
\label{p1p2} 
&&P_1(t)={1-e^{\beta \varepsilon}e^{-t/ \tau}  
\over 1+e^{\beta \varepsilon}}\nonumber\\  
&&P_2(t)={e^{\beta \varepsilon}(1+e^{-t/\tau})   
\over 1+e^{\beta \varepsilon}} \ .  
\end{eqnarray}  
The time-dependence of the system is thus characterized by an  
exponential function with a single relaxation time $\tau$  
that takes into account all possible probability fluxes   
\begin{eqnarray}  
&&\tau^{-1} \equiv W = \sum_{k,i \neq j} {1 \over \tau_{ij}^{k}}\nonumber\\  
&&=\tau_0^{-1} \left( e^{-\beta E_{\rm b}^{22}} +  
e^{-\beta E_{\rm b}^{12}} \right) 
\left( 1+e^{\beta \varepsilon}\right) \ .   
\end{eqnarray}  
As we see, $\tau$ is dominated by the lowest energy barrier $E_{\rm b}^{22}$, 
but with non-negligible pre-factors that take into account the  
possibility of recrossing from the equilibrium to the metastable state  
and the two different possibilities of jumping.  
Notice that this two prefactors are often neglected in theoretical  
studies of the dependence of the blocking temperature with the field \cite{Dormannadv97}  
and Monte Carlo simulations \cite{Lyberatosjpd00,Gonzalezprb96}. This is due to the fact that, usually, the 
possibility of jumping between minima by any of the two channels is not  
considered. However, at small non-zero fields ($\varepsilon \gtrsim 0$), and  
for particles oriented at $\psi \neq 0$, they can be equally relevant. 
This expression reduces to the usual one
\begin{equation}  
\tau^{-1}=\tau_0^{-1} e^{-\beta E_{\rm b}^{22}} 
\end{equation}  
when the energy function is symmetric ($\varepsilon=0$) and there is only  
one energy barrier, except for a factor 4 that can be absorbed in the definition of the  
prefactor $\tau_0$.  
  
The time dependence of the magnetization of the particle is then finally  
given by:   
\begin{eqnarray}  
&&m(t;K,\psi)=\cos[\theta_{\rm min}^1(\psi)]P_1(t)  
+\cos[\theta_{\rm min}^2(\psi)]P_2(t)\nonumber\\
&&  \nonumber\\   
&&=\bar{m}_{TS}(K,\psi)+ [m_0-\bar{m}_{TS}(K,\psi)]e^{-t/  
\tau(K,\psi)}.   
\end{eqnarray}  
In this equation, $\bar{m}_{TS}(K,\psi)$ is the equilibrium magnetization  
in the TSA [Eq. (\ref{magntsalf})], that has already been calculated in  
subsection \ref{2state}, and $m_0$ is the initial magnetization.  
If we have an ensemble of randomly oriented particles and a  
distribution of anisotropy constants $f(K)$, then the relaxation law 
of the magnetization is given by   
\begin{equation}  
\label{mtime} 
m(t)= \int _0 ^ \infty dK f(K) \int _0^ \pi d \psi  
\ m(t;K,\psi) \ .  
\end{equation}  
This will be the starting point for all the subsequent numerical calculations  
of the relaxation curves and magnetic viscosity.  
\section{Numerical calculations}  
\label{numerical}  
 
\subsection{Relaxation curves: $\svar$ scaling and normalization factors}  
 
In this section, we present the results of numerical calculations of  
the magnetization decay based on Eq. (\ref{mtime}) for a system of particles 
with logarithmic-linear distribution of anisotropy constants and random  
orientation of the easy-axis. For the sake of simplicity, we have assumed  
zero initial magnetization $m_{0}=0$, so particles have initially their  
magnetic moments at random and evolve towards the equilibrium state $m_{\rm eq}$.  
In the following, we will use dimensionless reduced variables for  
temperature and time, defined as $T/T_0$ and $t/\tau_0$, with  
$T_0=E_0/k_{\rm B}$ and $E_0$ the value of the energy at which $f(K)$ is centered. 
 
We have assumed that the magnetic moment of each particle is independent 
of the volume, although, in fact, it can be proportional to it, but  
this effect can be easily accounted by our model by simply changing  
$f(K)$ to $Kf(K)$ in all the expressions. For the case of a  
logarithmic-linear distribution, this change does not qualitatively modify  
the shape of the distribution. 
Other works \cite{Barbarajm93,Sampaiojm95,Marchandjm94} consider also a distribution 
of anisotropy fields $H_{\rm c}$ due to the spread of coercive fields in some 
real samples, but they study only relaxation rates at a fixed time.  
Here we have preferred to distribute $K$ and the easy-axes  
directions, which has a similar effect, in order to separate 
as much as possible the effects of an applied magnetic field from other 
effects that may possibly lead to non-conclusive interpretation of the 
results. 
 
\begin{figure}[htbp] 
\leavevmode 
\epsfxsize=19.2cm 
\epsfbox{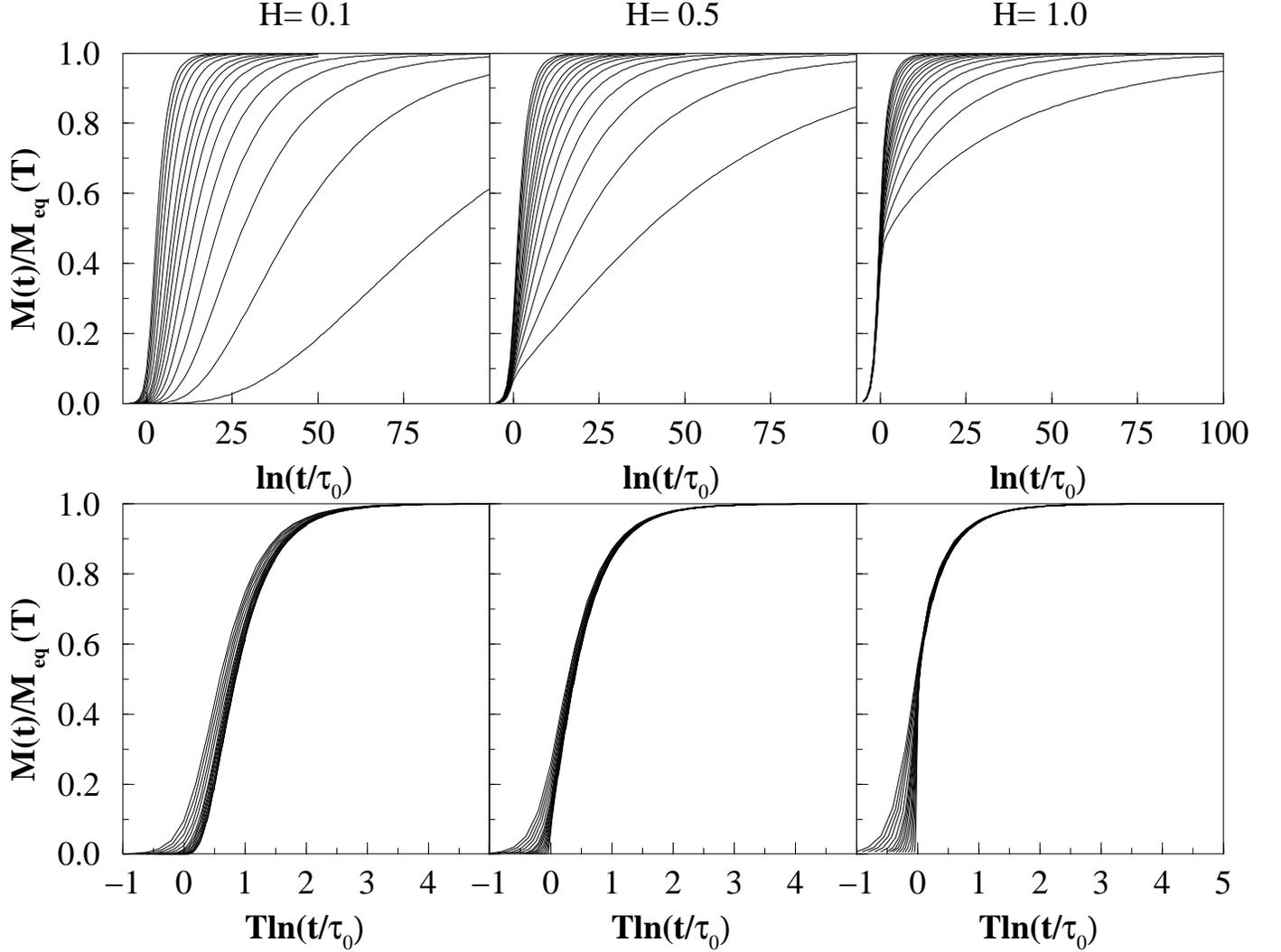} 
\caption{
Relaxation curves for an ensemble of particles with randomly oriented 
anisotropy axes and a logarithmic-normal distribution of anisotropies
$f(K)$ of width $\sigma=0.5$ and $K_0=1$ calculated by numerical integration of 
Eq. (\ref{mtime}). The initial magnetization has been set to $M_0=0$. 
Reduced temperatures $T/T_0$, starting from the lowermost curve, range from 
0.01 to 0.1 with 0.01 increments, and from 0.1 to 0.2 with 0.02 increments.
The applied fields are $H= 0.1, 0.5, 1.0$ as indicated.
The upper panels show the original curves normalized to the equilibrium 
magnetization $m_{TS}(T)$ given by Eq. (\ref{magnts}). 
In the lower panels, the same curves have been plotted as 
a function of the scaling variable $\svar$.
} 
\label{allrels05} 
\end{figure}

In Fig. \ref{allrels05}, we show the  
results of the numerical calculations for a system with $\sigma=0.5$ for three  
different fields $H=0.1, 0.5, 1.0$ and temperatures ranging from $0.02$ 
to $0.2$. In the upper panels, we present the original relaxations  
normalized to the equilibrium magnetization value as given by Eq. (\ref{magnts}).  
Normalization is essential in order to compare relaxations at different  
temperatures \cite{Balcellsprb97}, especially at low fields where the temperature dependence  
of the equilibrium magnetization is more pronounced.  
 
Our next goal is to investigate the possibility of scaling relaxation curves 
at different $T$ in a given magnetic field with the scaling variable $\svar$, 
in the spirit of our previous works \cite{Labartaprb93,Iglesiasjm95,Iglesiaszpb96,Balcellsprb97}. 
For this purpose, in the lower panels of Fig. \ref{allrels05},  
we show the relaxation curves of the upper panels as a function of the  
scaling variable $\svar$. 
According to Ref.\onlinecite{Labartaprb93}, in absence of a magnetic field,  
scaling should be valid up to temperatures such that $Te$ is of the order  
of $\sigma$. 
Instead, we observe in Fig. \ref{allrels05} 
that, the higher the field, the better the scaling of the curves is in 
the long time region and the worse at short times.
This observation holds independently of the value  
of $\sigma$, indicating that it is a consequence of the application of a 
magnetic field.   
This can be understood with the help of the effective energy barrier 
distribution introduced in Sec. \ref{effbar}. 
As was shown in Fig. \ref{distriali},
$h$ widens $f_{\rm eff}(E)$ and shifts the lowest energy barriers 
towards the origin, giving rise to a subdistribution of almost
zero energy barriers that narrows with $h$, and, consequently,
the requirements for $\svar$ scaling are worse fulfilled at small 
$\svar$ values.
On the contrary, as we will show in the next subsection, 
$h$ broadens the high energy tail of energy barriers that contribute 
to the relaxation, $f(E_{\rm b}^{22})$, improving the scaling requirements at
large $\svar$ values.    

\subsection{Scaling of relaxation curves at different magnetic fields} 
 
Another interesting point is the possibility of finding an appropriate 
scaling variable to scale relaxation curves at different fields for a given 
$T$, in a way similar to the case of a fixed field and different 
temperatures, in which $\svar$ is the appropriate scaling variable. 
In a first attempt, we will study the effect of $h$ on a system with random
anisotropy axes and the same $K=1$.  

\subsubsection{Randomly oriented particles, $K= 1$} 
 
We have calculated the relaxation curves for this system 
at $T=0.05$ and several values of the field. 
The obtained curves have been
normalized to the equilibrium magnetization as given by Eq. (\ref{magnts}). 
 
\begin{figure} 
\epsfxsize=8.6cm 
\leavevmode 
\epsfbox{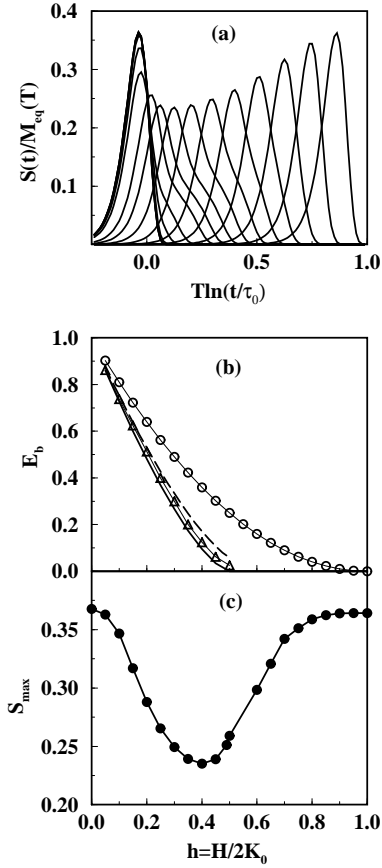}
\caption{(a) Low temperature ($T=0.05$) viscosity curves for a system of 
randomly oriented particles with the same anisotropy constant $K_0=1$.
The curves have been normalized to the equilibrium magnetization $M_{eq}(T)$
and correspond to magnetic fields $H= 0.1, 0.2,\dots , 1.0, 
1.2,1.4, 1.6, 1.8, 2.0$ increasing from right to left.
(b) Field dependence of the time corresponding to the maximum 
relaxation rate, $T\ln(t_{max}/\tau_0)$, as derived from the viscosity curves
in panel (a) (triangles). The field dependence of the mean lowest
energy barrier $E^{(2,2)}_{av}$ (diamonds), lowest energy barrier for 
particles oriented at $\alpha= 45^0$, $E^{(2,2)}(\pi/4)$ (squares) 
and  $\alpha= 0$ (circles), $E^{(2,2)}(0^0)$ 
are also shown for comparison. 
(c) Field dependence of the maximum relaxation rate $S_{max}$.
} 
\label{sindisrel} 
\end{figure}

The effect of $h$ on $M(t)$ is better understood in terms of  
the logarithmic time derivative of $M(t)$ 
\beq 
\label{St} 
S(t)= \frac{dM}{d(\ln(t))}= -\int_0^{\pi} d\psi  
\left(\frac{t}{\tau}\right)e^{-\frac{t}{\tau}} \ , 
\eeq 
which is the so-called magnetic viscosity $S(t)$. 
As can be clearly seen from Fig. \ref{sindisrel}a, the viscosity curves at  
different $h$ cannot be scaled neither by shifting them in the horizontal  
axis, nor by multiplicative factors, since the high and low field curves have  
different shapes. As soon as the field starts to destroy some of the energy  
barriers ($h\ge 0.5$), the qualitative form of the relaxation changes. This fact  
hinders, in principle, finding a field dependent scaling variable, valid in  
all the range of fields, in systems of non-aligned particles.  
 
Nevertheless, even though viscosity curves are qualitatively different at  
different $h$, all of them present a well-defined maximum corresponding to  
the inflection point of the relaxation curves.  
This maximum appears at a time  
$t_{\rm max}$ associated to an $E_{\rm max}= T\ln(t_{\rm max} /\tau_0)$, 
that decreases with increasing $h$ for a given temperature.  
This energy is approximately equal to the averaged  
lowest energy barrier of the particles ($E_{\rm b}^{22}$ in the notation of Sec.  
\ref{model}) and this value is closer to the lowest possible barrier  
(corresponding to particles oriented at $\psi= 45^{\circ}$) than to the  
barrier of a particle aligned with the field.  
In Fig. \ref{sindisrel}b, we have plotted the field dependence of all  
this quantities, together with  
the position of the maximum of the viscosity in energy units  
$E_{\rm max}$, and in Fig. \ref{sindisrel}c the value of the  
corresponding viscosity $S_{\rm max}$. 
 
The reduction of $t_{\rm max}$ with $h$ can be understood in terms of the  
progressive reduction of the energy barriers by $h$.  
At $h=0$, the barriers are independent of the orientation of the particle and  
equal to 1, so that the maximum is placed at $E_{\rm max}=1$ and $S_{\rm max}=1/e$  
according to Eq. (\ref{St}).  

For $h\ge0.5$ (the critical field for particles oriented
at $\psi=45^{\circ}$), the lowest energy barriers start to be destroyed by $h$
and consequently the relaxation rates peak at $E_{\rm max}=0$ with an increasing
$S_{\rm max}$ value that increases as more particles loose their barriers.
For $h\ge 1$ all barriers have been  
destroyed and relaxations become field independent, with $E_{\rm max}=0$ and  
$S_{\rm max}=1/e$.  
For fields up to $h=0.5$, the variation of $E_{\rm max}$ and  
$S_{\rm max} $with $h$ can be used to scale the magnetic relaxation
curves at constant $T$ and different $h$. 
Therefore, although in this case
the inflection points of the relaxation curves could be brought together
by shifting them in the $\svar$ axis in accordance with the variation of 
$E_{av}^{22}$, the full scaling cannot be accomplished because of 
the complicated variation of $S_{\rm max}$ (see Fig. \ref{sindisrel}c).  
 
\subsubsection{Randomly oriented particles with $f(K)$} 
 
In spite of the lack of scaling of the preceding case, in what follows, 
we will demonstrate that the inclusion of a distribution of $K$, 
always present in experimental systems, allows to
scale the relaxation curves for a wide range of $h$.

\begin{figure} 
\epsfxsize=8.6cm 
\leavevmode 
\epsfbox{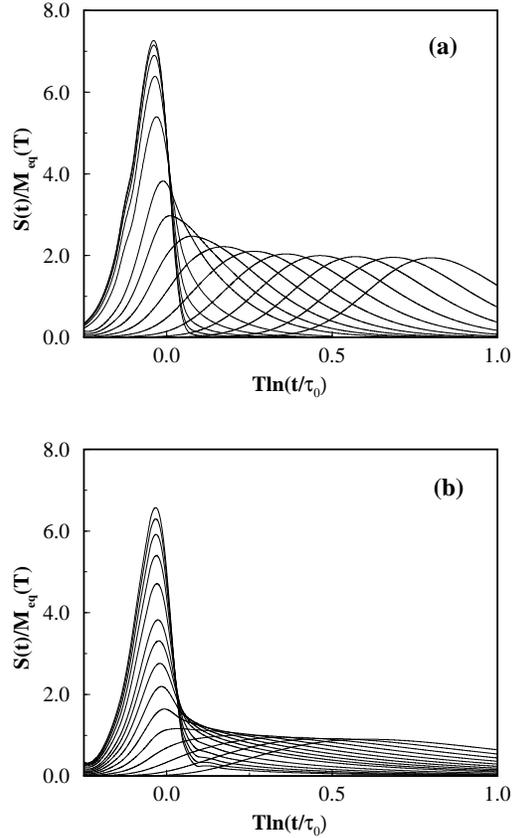}
\caption{(a) Low temperature ($T=0.05$) viscosity curves for a system of 
particles with random orientations and logarithmic-normal distribution 
of anisotropies with (a) $\sigma= 0.2$ and (b) $\sigma= 0.5$.
The curves have been normalized to the equilibrium magnetization and 
correspond to magnetic fields $H= 0.1$ to $1.0$ in 0.1 steps and $H=1.2$
to $2.0$ in 0.2 steps starting from the right.
} 
\label{sig02t_1.ali} 
\end{figure}

Let us consider a logarithmic-linear distribution of anisotropy constants of  
width $\sigma$, Eq. (\ref{logn}). Low temperatures relaxation rates 
corresponding to $\sigma= 0.2, 0.5$ are presented in Fig. \ref{sig02t_1.ali}.  
In this case, the qualitative shape of the viscosity curves is not distorted
by $h$. It simply shifts the position of the maxima towards lower values 
of $\svar$ and narrows the width of the peaks, being these effects similar
for both studied $\sigma$. 
 
\begin{figure} 
\epsfxsize=8.6cm 
\leavevmode 
\epsfbox{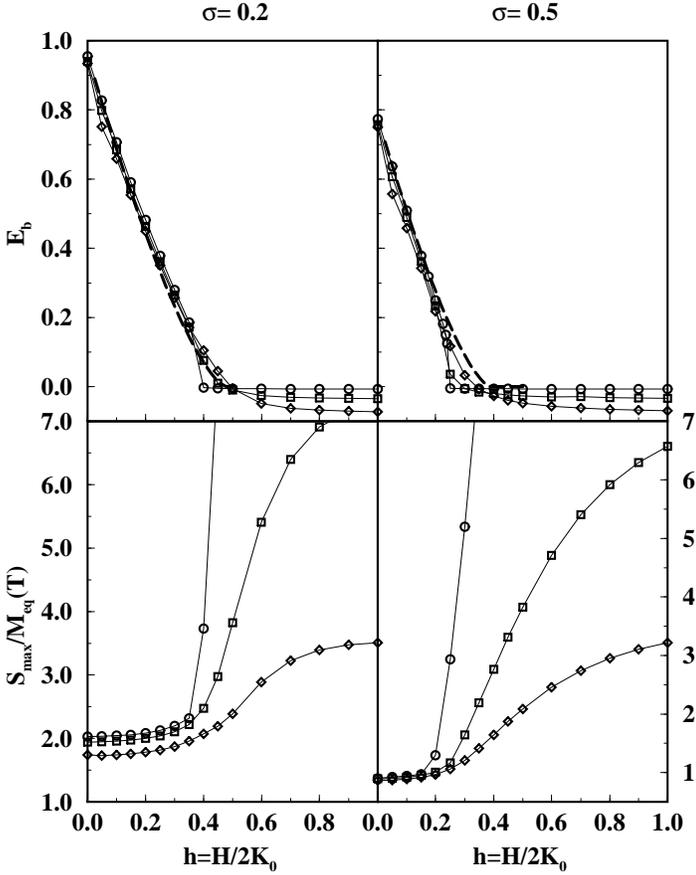}
\caption{Upper panels: Field dependence of the energy corresponding to the maximum 
relaxation rate, $T\ln(t_{max}/\tau_0)$, as derived from the viscosity curves
in Fig. \ref{sig02t_1.ali} for temperatures $T=0.01$ (circles), 0.05 (squares)
, 0.1 (diamonds). 
Lower panels: Field dependence of the maximum relaxation rate $S_{max}$ for the same curves and temperatures. Left column is for $\sigma= 0.2$ and the right one for $\sigma=0.5$.
} 
\label{smaxsig02.ali} 
\end{figure}

The position of the maximum relaxation rate still decreases with increasing  
$h$, following the decrease of the smallest energy barriers
(see Fig. \ref{smaxsig02.ali}), which now have an almost linear dependency
on $h$. As in the preceding case, $E_{\rm max}$ goes to zero when $h$ starts
to destroy the lowest energy barriers.  
The difference is now that, due to the spread of the anisotropy constants,  
lower fields $h_0$ are needed to start extinguishing the lowest energy barriers 
(see Fig. \ref{smaxsig02.ali}), being this reduction greater, the greater
$\sigma$ is, since the most probable anisotropy constant 
[$K_{\rm max}$=$K_0\exp(-\sigma^2/2)$] becomes smaller 
and, consequently, $E_{\rm max}$ drops to zero at smaller $h_0$.
This field corresponds to the one for which $f_{\rm eff}(E_{\rm b}^{22})$
starts to develop a peak corresponding to zero energy barriers.
As in the preceding case, we have also tried to identify the variation 
of $E_{\rm max}$ with the microscopic energy barriers of the system.
As can be clearly seen in the dashed lines of the upper panels of Fig.
\ref{smaxsig02.ali}, the $h$ dependence of $E_{\rm max}$ follows that
of the lowest energy barriers for particles oriented at $\psi= \pi/4$
and with $K=K_{\rm max}$.

By looking in detail the low $T$ relaxation curves 
($T=0.01$ curves, analogue to the ones shown in Fig. \ref{sig02t_1.ali} for $T= 0.05$),
we have observed that two relaxation regimes can be distinguished.
One presents a broad peak in $S$ at relatively high energies (long times)
with a maximum at an $E_{\rm max}$ which varies as the lowest energy
barrier at $\psi= 45$, this is clearly visible at the lowest $h$ values
even for the $T=0.05$ case of Fig. \ref{sig02t_1.ali}.
The other regime presents a peak around $E=0$ that starts to develop as 
soon as $h$ breaks the lowest energy barriers. 
What happens is that the first peak shifts towards lower energies
with $h$, at the same time that the relative contribution of the second peak
increases, the global effect being that, at a certain $h$, the contribution
of the first peak has been swallowed by the second because
at high $h$ and low $T$, relaxation is driven by almost 
zero energy barriers.

\begin{figure} 
\epsfxsize=8.6cm 
\leavevmode 
\epsfbox{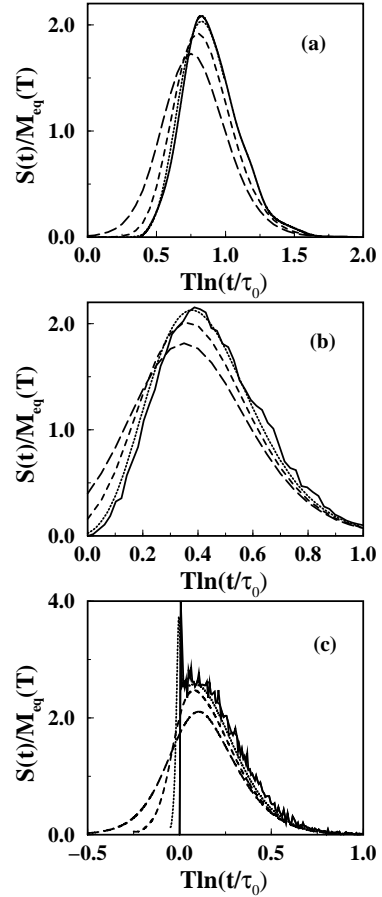}
\caption{Relaxation rates as a function of the scaling variable $\svar$ for 
different temperatures 
[$T= 0.01$ (dotted line), 0.05 (dashed line) , 0.1 (long-dashed line)], 
$\sigma= 0.2$ and three magnetic fields 
(a) $H= 0.1$, (b) $H= 0.5$, and (c) $H= 0.8$. 
The curves tend to the effective distribution of lower energy barriers 
$f_{\rm eff}(E^{22})$, shown as a continuous line, as $T$ decreases.} 
\label{visc+diss02.ali} 
\end{figure}

To clarify this point, we show in Fig. \ref{visc+diss02.ali} $S(t)/T$  
for three different temperatures and  
magnetic fields $H= 0.1, 0.5, 1.0$ for a narrow ($\sigma= 0.2$) and a wide  
($\sigma= 0.5$) distribution. The effective distribution of lowest energy  
barriers $f_{\rm eff}(E^{22})$, already calculated in Fig. \ref{distriali}, is also  
plotted as a continuous line.  
We observe that for a narrow distribution, at low enough $T$,  
$S(t)/T$ coincides with $f_{\rm eff}(E^{22})$  
independently of $h$, demonstrating that only the lowest energy barriers  
of the system contribute to the relaxation. 

\begin{figure} 
\epsfxsize=8.6cm 
\leavevmode 
\epsfbox{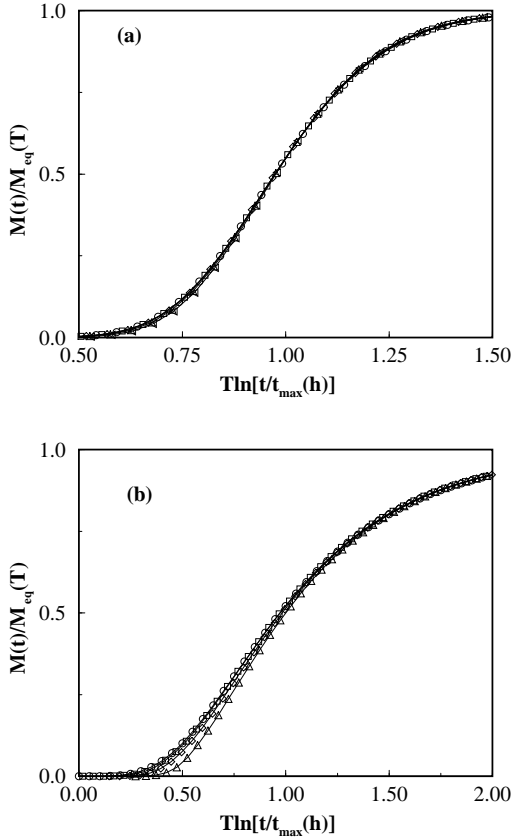}
\caption{Normalized relaxation curves as a function of the scaling variable 
$T\ln[(t/t_{max}(h)]$ for $T= 0.05$) obtained from Fig. \ref{sig02t_1.ali}
by shifting the curves in the horizontal axis with the position of 
the maximum relaxation rate (upper panels in Fig. \ref{smaxsig02.ali}).
(a) $\sigma= 0.2$ and $H= 0.1, 0.2, 0.3, 0.4, 0.5$; 
(b) $\sigma= 0.5$ and $H= 0.1, 0.2, 0.3, 0.4$ 
(starting from the uppermost curve)}. 
\label{fig13} 
\end{figure}

Finally, let us also notice that, at difference with the preceding case, $S_{\rm max}$ becomes almost constant below $h_0$ (lower panels in Fig. \ref{smaxsig02.ali}) and low enough $T$, so that now the relaxation curves 
at different $h$ and fixed $T$ may be brought to a single curve
by shifting them along the $\svar$ axis in accordance to the $E_{\rm max}$ 
variation. 
The resulting curves are displayed in Fig. \ref{fig13}
for $\sigma = 0.2, 0.5$. They are the equivalent of the master curves of
Fig. \ref{allrels05} for a fixed $h$ and different $T$. 
Now the appropriate scaling variable is
\begin{equation}
E_{\rm sca}=T\ln[t/t_{\rm max}(h)]	\ ,
\end{equation}
which generalizes the scaling at fixed $T$.
This new scaling is valid for fields lower than $h_0$, the field at which the
lowest barriers start to be destroyed and above which the relaxation becomes dominated by almost zero energy barriers. Thus, as already discussed in the previous paragraphs, the wider $\sigma$, the smaller the $h$ range for the validity of field scaling.
 
\section{conclusions}  
\label{conclusions}  

We have proposed a model for the relaxation of small particles systems under a magnetic field which can be solved analytically and which allows to study the effect of the magnetic field on the energy barrier distribution. In particular, we have shown that the original $f(E_{\rm b})$ is splitted into two subdistributions which evolve towards higher and lower energy values, respectively, as $h$ increases. It is precisely the  
subdistribution of lowest energy barriers, the one that completely dominates the relaxation as it is evidenced by its coincidence with the relaxation rate
at low $T$. 

For fields smaller than the critical values for the smallest barriers,
the relaxation curves at different $h$ and fixed $T$ can be collapsed into
a single curve, in a similar way than $\svar$ scaling for curves at fixed $h$.
Whereas the latter allows to extract the barrier distribution by differentiation of the master curve \cite{Balcellsprb97}, the shifts in the $\svar$
axis necessary to produce field scaling, give the field dependence of the mean
relaxing barriers, a microscopic information which cannot easily be inferred from other methods \cite{Sappeyprb97}.   
\section*{Acknowledgements}
We acknowledge CESCA and CEPBA under coordination of 
C$^4$ for the computer facilities. This work has been supported by 
SEEUID through project MAT2000-0858 and CIRIT under project 2000SGR00025.
  


\end{multicols} 
\twocolumn 
 
\newpage
\onecolumn
\twocolumn

\end{document}